\documentclass[aps,prc,twocolumn,showpacs,superscriptaddress,longbibliography,floatfix,10pt]{revtex4-1}
\usepackage{bm}
\usepackage{amssymb}
\usepackage{amsmath}
\usepackage{graphicx}
\usepackage{xcolor}
\usepackage[colorlinks,linkcolor=blue]{hyperref}
\begin{document}
\title{Swelling of nuclei embedded in neutron-gas and consequences for fusion}

\author{A.S. Umar}\email{umar@compsci.cas.vanderbilt.edu}
\author{V.E. Oberacker}\email{volker.e.oberacker@vanderbilt.edu}
\affiliation{Department of Physics and Astronomy, Vanderbilt University, Nashville, TN 37235, USA}
\author{C. J. Horowitz}\email{horowit@indiana.edu}
\affiliation{Department of Physics and CEEM,
             Indiana University, Bloomington, IN 47405, USA}
\author{P.-G. Reinhard}\email{paul-gerhard.reinhard@physik.uni-erlangen.de}
\affiliation{Institut f\"ur Theoretische Physik, Universit\"at Erlangen, D-91054 Erlangen, Germany}
\author{J.A. Maruhn}\email{maruhn@th.physik.uni-frankfurt.de}
\affiliation{Institut f\"ur Theoretische Physik, Goethe-Universit\"at, D-60438 Frankfurt am Main, Germany}
\date{\today}
\begin{abstract}
Fusion of very neutron rich nuclei may be important to determine the composition and heating of the
crust of accreting neutron stars.
We present an exploratory study of the effect of the neutron-gas environment on the structure of 
nuclei and the
consequences for pycnonuclear fusion cross-sections in the neutron drip region. 
We studied the formation and properties of
Oxygen and Calcium isotopes embedded in varying neutron-gas densities. We observe
that the formed isotope is the drip-line nucleus for the given effective interaction.
Increasing the neutron-gas density leads to the swelling of the nuclear density.
We have used these densities to study the effect of this swelling on the fusion
cross-sections using the S\~{a}o-Paulo potential. At high neutron-gas densities the
cross-section is substantially increased but at lower densities the modification is
minimal.
\end{abstract}
\smallskip
\pacs{25.60.Pj, 
26.60.+Gj, 
97.80.Jp 
}
\maketitle
\raggedbottom
\section{Introduction}
Recent advances in radioactive beam technologies have opened up new experimental possibilities to
study fusion of neutron rich nuclei~\cite{balantekin2014}.
Furthermore, near barrier fusion cross sections are relatively
large so experiments are feasible with modest beam intensities.
In addition, measurements are possible at the TRIUMF ISAC facility and in the near future at the
NSCL ReA3-6 reaccelerated beam facility. Other radioactive ion beam facilities include
ATLAS-CARIBU at Argonne National Laboratory, SPIRAL2 at GANIL (France), and RIBF at RIKEN (Japan).
Note that the dynamics of the neutron rich skin of these nuclei can enhance the cross-section over
that predicted by a simple static barrier penetration model. For example, neutrons may be transferred
from the neutron rich beam to the stable target. Fusion of very neutron rich nuclei, near the drip
line, raise very interesting nuclear structure and nuclear dynamics questions.

Neutron stars, in binary systems, can accrete material from their companions.
This material undergoes a variety of nuclear reactions~\cite{haensel2007}.  
First at low densities, conventional
thermonuclear fusion takes place, see for example~\cite{schatz2001}.
Next at higher densities, the rising electron Fermi energy induces a series of electron
captures~\cite{gupta2007} to produce increasingly neutron rich nuclei. Finally at high densities,
these very neutron rich
nuclei can fuse via pycnonuclear reactions. Pycnonuclear fusion is induced by quantum zero point
motion~\cite{salpeter1969,schramm1990}. The energy released, and the densities at
which these reactions occur, are
important for determining the temperature and composition profile of accreting neutron star crusts.
The existence of the inner neutron-star crust, in which very neutron rich nuclei are immersed in a 
gas of neutrons raises the question, what is the impact of this neutron gas on nuclear fusion rates?
This neutron drip region is believed to occur for densities 
in the approximate range of  a few $\times10^{11}$ to $8\times10^{13}$ g/cm$^3$.
One of the early studies of the inner crust, consisting of drip-line nuclei combined with the background
neutron gas, had been done by Negele and Vautherin~\cite{negele1973}.
Therefore understanding fusion reactions of neutron rich isotopes near the drip
line are important.
Horowitz \textit{et al.}~\cite{horowitz2008} calculate the enhancement in fusion rates from strong ion 
screening using molecular
dynamics simulations, and find that $^{24}$O + $^{24}$O can fuse near
$10^{11}$~g/cm$^3$, just before neutron drip. 
Extensive studies of the astrophysical $S(E)$ factors have been
done using densities emanating from microscopic calculations and a barrier penetration model
for fusion~\cite{beard2010,afanasjev2012}.
Furthermore, this fusion can take place in the background neutron gas that is
present in the inner crust of a neutron star.
The possible effect of the neutron gas background was discussed in Ref.~\cite{afanasjev2012}
by empirically changing the barrier height and width.
Here, we study this effect by considering the presence of the background neutron gas
microscopically by directly including the neutron gas and the nucleus in the same framework. 
We explicitly calculate the self-consistent proton and neutron densities of a single nucleus 
in equilibrium with the background neutron gas.  Furthermore, for astrophysical applications, 
it seems clear that this adiabatic approach is the one that is
relevant for calculating fusion rates in the inner crust. The neutron gas should have plenty
of time to adjust to the presence of a nucleus.

The paper is organized as follows. Our computational approach to consider the presence of
neutron gas together with the nucleus and the model for calculating the fusion cross-sections
is discussed in Sec.~\ref{sec2}. Computational results for the Oxygen and Calcium systems is
described in Sec.~\ref{sec3}. Finally, these
results are discussed and we conclude in Sec.~\ref{sec4}.

\section{Computational Details\label{sec2}}
\label{sec.formalism}
\subsection{Computational setup}
Hartree-Fock (HF) calculations were done in a three-dimensional Cartesian geometry with no symmetry assumptions
and using the Skyrme effective nucleon-nucleon interaction~\cite{umar2006c}.
The infinite neutron star crust environment is simulated by using a three-dimensional Cartesian
box with periodic boundary conditions for both the bound and neutron gas states as well as the solution
of the Poisson equation for the Coulomb potential, which is performed using Fast-Fourier Transform (FFT)
techniques~\cite{maruhn2014}.
The Coulomb solution assumes global neutrality which means that the proton charges are compensated by
a homogeneous negative electron gas cloud. In practice, this is achieved by setting the zero-momentum
part of the Coulomb field in Fourier space to zero.
The code uses the basis-spline collocation method
for the lattice discretization of the HF equations using periodic boundary-conditions as described in
Refs.~\cite{umar1991a,umar1991b,bottcher1989}. The HF equations are solved using the damped gradient iteration method.
The Skyrme parametrization used was SLy4~\cite{chabanat1998a}. In addition to providing a good description
of nuclei
this interaction has been used to produce an equation of state for neutron stars~\cite{douchin2001}.

For the choice of initial states to be used in HF minimization we have tried a number of choices, which
all resulted in the same identical solution. One can first generate any isotope of the desired nucleus by solving the
HF equations as described above and subsequently combine these states with a large number of free neutron
gas states and minimize the entire system again. Alternately, one can simply choose a number of free
proton states together with a large number of free neutron states and minimize this system.
Both methods result in exactly the same numerical solution with a drip-line isotope corresponding
to the nucleus with the given number of protons embedded in a given density of neutron gas states.
Initial states, $\boldsymbol{\psi}$, are spinors with a non-zero upper component in case of
time-reversal invariance. In case of no time-reversal invariance the number of states are doubled by
adding spinors having non-zero lower components as well. They satisfy the periodicity condition
\begin{equation}
 \boldsymbol{\psi}_{\mathbf{n}}(\mathbf{r}+\mathbf{L})=\boldsymbol{\psi}_{\mathbf{n}}(\mathbf{r})\;,
\end{equation}
where $\mathbf{n}=(n_x,n_y,n_z)$ and $n_a$ taking on integer values $-N_a,\ldots,+N_a$.
Free states satisfying the above periodicity condition are simple plane-wave states with the
appropriate normalization
\begin{equation}
 \boldsymbol{\psi}_{\mathbf{n}}(\mathbf{r})=\frac{1}{\sqrt{L_xL_yL_z}}e^{\imath(k_{n_x}x+k_{n_y}y+k_{n_z}z)}
 \boldsymbol{\chi}_n\;,
\end{equation}
where $k_{n_a}=2\pi n_a/L_a$ and $\boldsymbol{\chi}_n$ is an up or down spinor.
The initial neutron and proton densities are perfectly uniform filling the
entire numerical box.
These initial states comprise the total number of states used in the self-consistent HF problem using
the Skyrme interaction. For even number of states time-reversal is valid and the HF single-particle
Hamiltonian only depends on particle density, $\rho$, kinetic energy density. $\tau$, and the spin-orbit
pseudotensor $\mathbf{J}$ through the single-particle states~\cite{chabanat1998a}
\begin{equation}
\mathbf{h}\left( \left\{ \boldsymbol{\phi}_{\mu} \right\} \right) \boldsymbol{\phi}_{\lambda}=\epsilon_{\lambda}\boldsymbol{\phi}_{\lambda}
\;\;\;\;\;\;\;\;\;\lambda=1,...,N\;.
\label{tdhf0}
\end{equation}
As the HF iterations proceed (preserving orthogonality for the entire system)
some states evolve to form a bound nuclear system while the others remain as gas states showing
some non-uniformity due to the presence of shell effects.

\subsection{Fusion cross-sections}

The S\~{a}o Paulo model of fusion calculates an effective nuclear potential based on the density overlap
between colliding nuclei
\cite{gasques2004,chamon2002}. Sub-barrier fusion cross-sections can then be calculated via tunneling.
The model can be easily applied to
a very large range of fusion reactions and qualitatively reproduces many experimental
cross-sections~\cite{gasques2005,gasques2007}.
Recently this model was used to tabulate astrophysical $S$ factors describing fusion of many carbon, oxygen, neon and
magnesium isotopes for use in astrophysical simulations~\cite{beard2010}, see also Ref.~\cite{yakovlev2010}.

In this section we describe the S\~{a}o Paulo barrier penetration model to calculate fusion cross-sections. This
starts with the double folding potential $V_F(R)$~\cite{gasques2004,chamon2002},
\begin{equation}
V_F(R)=\int d^3r_1d^3r_2 \rho_1(r_1)\rho_2(r_2) V_0 \delta(\mathbf{r}_1- \mathbf{r}_2 - \mathbf{R})\, .
\label{eq:vf}
\end{equation}
Here $\rho_1$ and $\rho_2$ are the densities of the two nuclei and $V_0=-450$ MeV-fm$^{3}$.
From $V_F$ a nonlocal potential $V_N(R,E)$ is constructed, $V_N(R,E)=V_F(R)e^{-4v^2/c^2}$, where $v$ is the local relative
velocity~\cite{gasques2004,chamon2002} between the two nuclei at separation $R$ ($c$ is the speed of light)
\begin{equation}
 v^2(R,E)=\frac{2}{\mu}\left[E-V_C(R)-V_N(R,E)\right]\;.
 \label{eq:v2}
\end{equation}
Here $\mu$ is the reduced mass and $V_C(R)$ is the Coulomb potential at $R$.
In practice, we use FFT techniques to calculate $V_F(R)$ as well as the Coulomb
potential $V_C(R)$ (instead of using the point Coulomb formula). The velocity
equation~(\ref{eq:v2}) has to be solved by iteration at each value of $R$ and $E$.  

Note that the neutron gas background could behave differently for two nuclei in close proximity 
then it does for only a single nucleus. However for simplicity, in this first study, we consider 
only a single nucleus in the background gas at a time in order to get the density profiles 
shown in Fig. \ref{fig2}.  We then use these profiles in Eq. \ref{eq:vf} and assume they 
are unmodified by the presence of the second nucleus.

The fusion barrier penetrabilities $T_L(E_{\mathrm{c.m.}})$
are obtained by numerical integration of the two-body Schr\"odinger equation
\begin{equation}
\left[ \frac{-\hbar^2}{2\mu}\frac{d^2}{dR^2}+\frac{L(L+1)\hbar^2}{2\mu R^2}+V(R,E)-E\right]\psi=0\;,
\label{eq:xfus}
\end{equation}
using the {\it incoming wave boundary condition} (IWBC) method~\cite{hagino1999}.
The potential $V(R,E)$ is the sum of nuclear and Coulomb potentials.
IWBC assumes that once the minimum of the potential is reached fusion will
occur. In practice, the Schr\"odinger equation is integrated from the potential
minimum, $R_\mathrm{min}$, where only an incoming wave is assumed, to a large asymptotic distance,
where it is matched to incoming and outgoing Coulomb wavefunctions. The barrier
penetration factor, $T_L(E_{\mathrm{c.m.}})$ is the ratio of the
incoming flux at $R_\mathrm{min}$ to the incoming Coulomb flux at large distance.
Here, we implement the IWBC method exactly as it is
formulated for the coupled-channel code CCFULL described in Ref.~\cite{hagino1999}.
This gives us a consistent way for calculating cross-sections at above and below
the barrier via
\begin{equation}
\sigma_f(E_{\mathrm{c.m.}}) = \frac{\pi}{k^2} \sum_{L=0}^{\infty} (2L+1) T_L(E_{\mathrm{c.m.}})\;.
\label{eq:sigfus}
\end{equation}

\section{Results\label{sec3}}
All the calculations presented here were done using a three-dimensional cubic Cartesian box with
$31$~fm sides and $1.0$~fm lattice spacing. With the basis-spline method this gives highly accurate
results for the HF problem~\cite{umar1991a}. We studied two systems with $Z=8$ and $Z=20$ protons
inside a neutron gas. The resulting nuclei are always spherical and density profiles can be
obtained by taking a cut along a particular axis. Our mesh includes the origin in all directions.
We have repeated some of these calculations in a cubic bot with $41$~fm sides and for the same
neutron-gas density the results were numerically indistinguishable.
\begin{figure}[!htb]
	\includegraphics*[width=8.6cm]{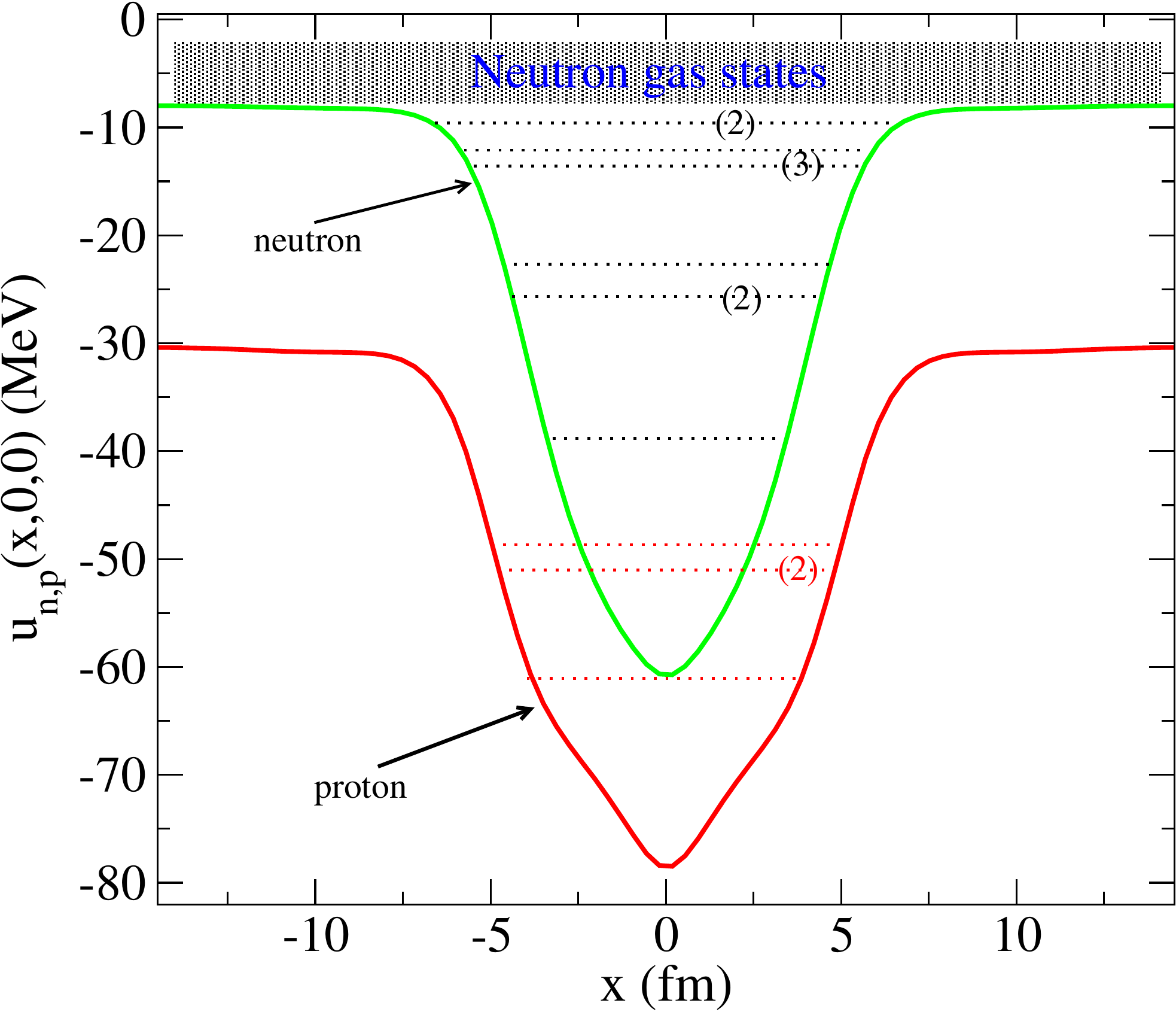}
	\caption{\protect (Color online) Mean-field potentials for neutrons and protons (solid lines) for the
		system with $Z=8$ and $520$ neutrons. The dashed lines indicate the energies of the bound single-particle
		states with degeneracies shown in brackets.}
	\label{fig1}
\end{figure}

\subsection{$Z=8$ system}

In these calculations we start with $8$ proton states and a number of neutron states
ranging from $50-1020$. The Skyrme SLy4 force gives $^{28}$O as the slightly bound
drip-line nucleus in free-space. For the above range of neutron states we also find the
$^{28}$O to be the bound part of the system. As the number of neutron states is increased
an overall negative potential is developed permeating the entire box.
Figure~\ref{fig1} shows the neutron and proton mean-field potentials (solid lines) for
$Z=8$ and $520$ neutrons. The dashed lines indicate the energies of the bound single-particle
states with degeneracies shown in brackets.
We define $N_\mathrm{bound}$
as the highest neutron s.p. state below the continuum threshold.
Consequently, the bound and gas densites become
\begin{eqnarray}
\rho_\mathrm{bound}
&=&
\sum_{\lambda=1}^{N_\mathrm{bound}}
\left|\phi_\lambda\right|^2
\; ,
\\
\rho_\mathrm{free}
&=&
\sum_{\lambda>N_\mathrm{bound}}
\left|\phi_\lambda\right|^2 \; .
\end{eqnarray}

\begin{figure}[!htb]
	\includegraphics*[width=8.6cm]{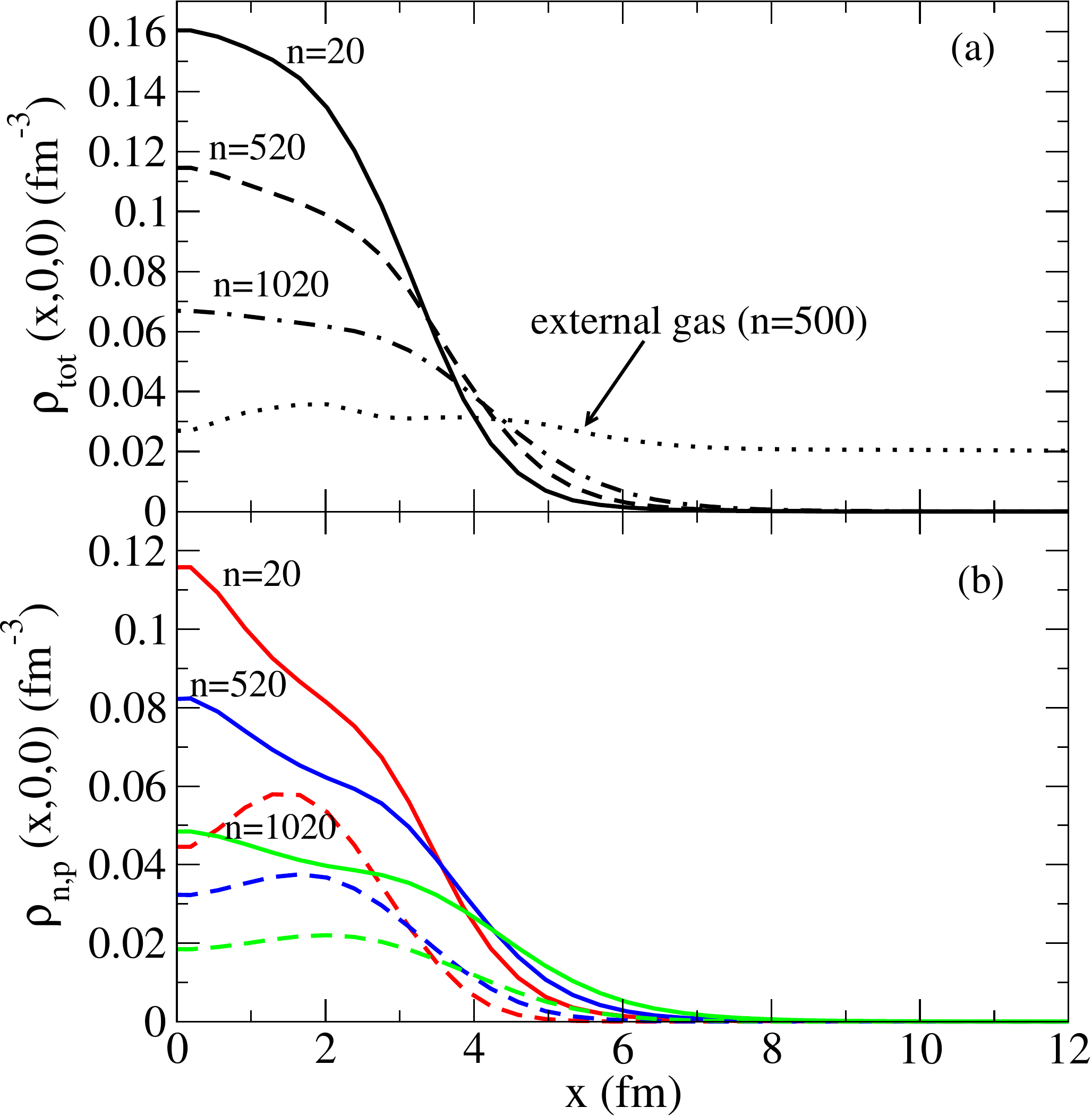}
	\caption{\protect (Color online) (a) Total density profiles for bound states; (b) density profiles for bound neutrons
		(solid lines) and protons (dashed lines), for the system with $Z=8$ and $n=20,520$, and $1020$ neutrons.}
	\label{fig2}
\end{figure}
In this case the
asymptotic value of the neutron potential is about $-8$~MeV.
As it is the case in free-space increasing neutron number leads to deepening of the proton potential.
In Fig.~\ref{fig2} we show the density profiles for neutrons and protons as well as the total density
as a function of the number of neutron-gas states. The top frame shows the total density behavior
as the neutron-gas density is increased. The curves labeled $n=20$ correspond to free-space $^{28}$O
nucleus. As the external neutron-gas density is increased the bound system swells
up in a way similar to a density scaling $\rho(r)\rightarrow\rho(sr)$ with $s<1$ as discussed
in Ref.~\cite{umar2007a}.
While the peak of the total density decreases from the free-space value of $0.16$~fm$^{-3}$ to
as low as $0.068$~fm$^{-3}$ for the $1000$ external neutron-gas state case, the tail region flattens
and develops a larger spatial extent, since the total integral remains to be $28$.
The density profiles are symmetric about $x=0$ and the numerical box extends to larger values then
shown in the figure.
\begin{figure}[!htb]
	\includegraphics*[width=8.6cm]{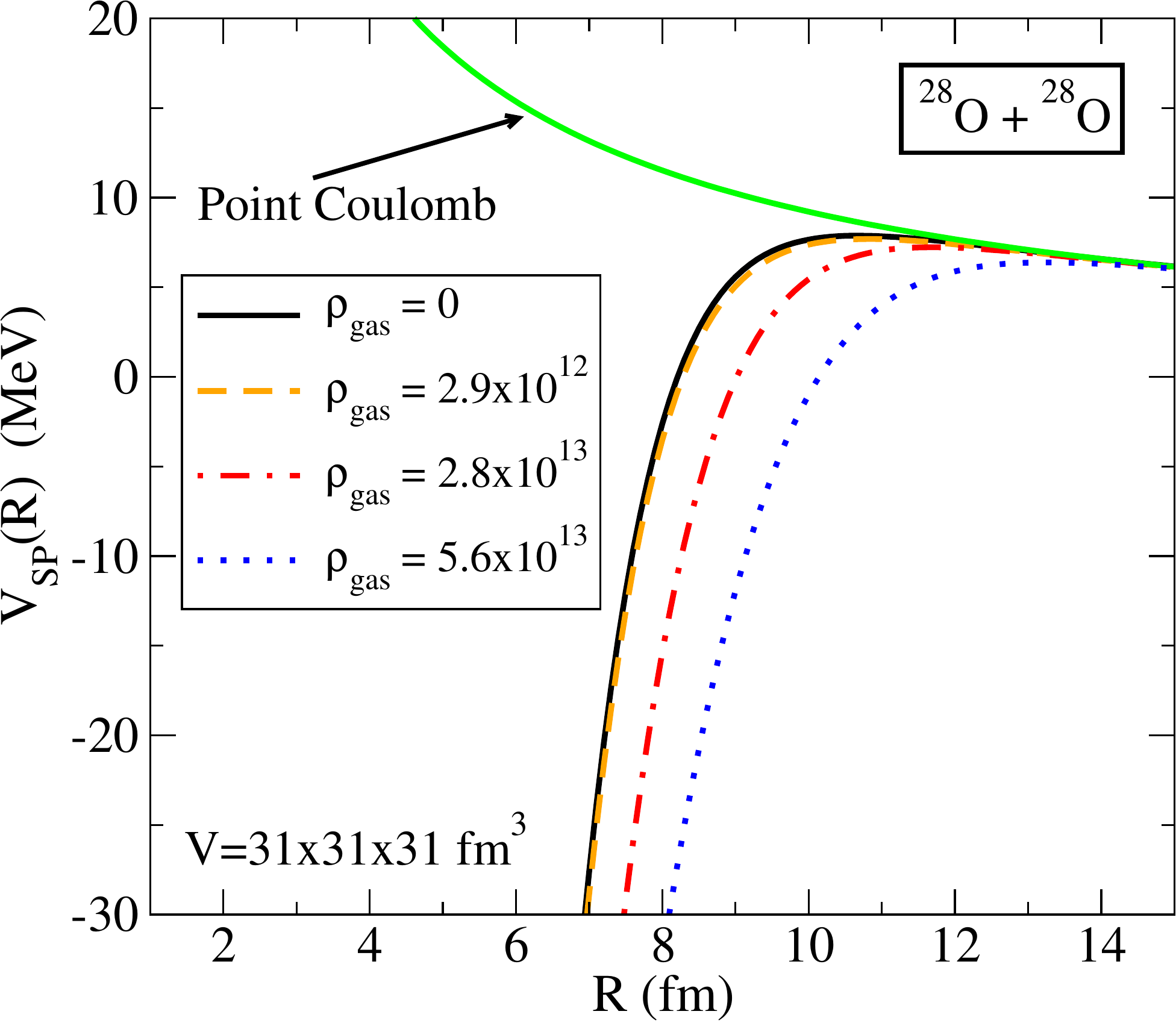}
	\caption{\protect (Color online) Ion-Ion potentials $V_{SP}(R)=V_F(R)+V_C(R)$ for $^{28}$O isotope
		as a function of the external neutron-gas density. Also shown is the point-Coulomb interaction.
		Densities are in units of gm/cm$^3$.}
	\label{fig3}
\end{figure}
Using these densities in Eq.~(\ref{eq:vf}) we have calculated the corresponding ion-ion folding
potentials as well as the Coulomb potential.
In Fig.~\ref{fig3} these potentials are plotted for a range of external neutron-gas densities.
What is observed is that for neutron-gas densities in
the range $\rho_{gas}=2-4\times 10^{12}$~gm/cm$^3$ the effect of the gas is not changing the
ion-ion potential in comparison to the free-space case in a considerable way. 
However, for gas densities
above $10^{13}$~gm/cm$^3$ a very significant change is observed.
\begin{figure}[!htb]
	\includegraphics*[width=8.6cm]{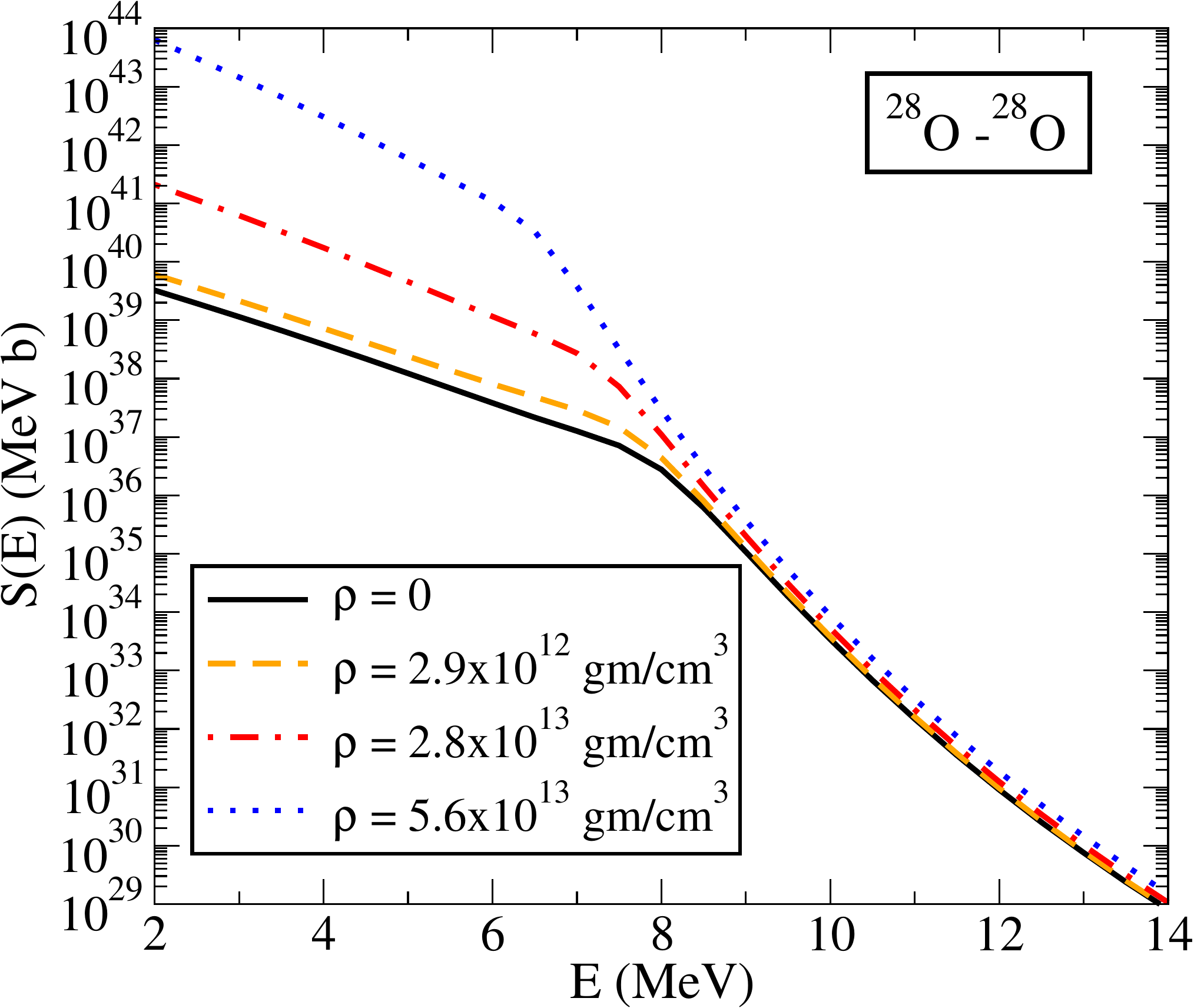}
	\caption{\protect (Color online) Astrophysical $S$ factor versus center of mass energy for fusion of $^{28}$O isotope
		as a function of external neutron-gas density. Cross-sections are calculated using the S\~{a}o Paulo barriers and
		the IWBC method.}
	\label{fig4}
\end{figure}
The free-space barrier has a peak value of $7.87$~MeV located at $R=10.7$~fm. As the external gas
density is increased the corresponding barrier height is reduced to $7.69$, $7.23$, and $6.39$~MeV with the peak
location moving outward at $10.8$, $11.7$, and $13.3$~fm.
This behavior is very similar to what is observed in free-space as one goes up in the oxygen isotope
chain~\cite{umar2012a}.
Unfortunately, the dynamical density constrained time-dependent Hartree-Fock 
(DC-TDHF) methodi~\cite{umar2006b,umar2008a,keser2012} used in 
Ref.~\cite{umar2012a} is not applicable in this situation since it requires
a fully dynamical calculation.

These together with the calculation of the energy
dependence from Eq.~(\ref{eq:v2}) allows the calculation of fusion cross-sections as a function of external neutron-gas
density, using the IWBC method discussed in the previous section. Figure~\ref{fig4} demonstrates the effect
of neutron-gas seen in the potentials on the astrophysical $S$ factor.
The S-factor for different external neutron-gas densities start to deviate from each other as soon as
the center-of-mass energy falls below the barrier. Even for the lowest gas density of $2.9\times 10^{12}$~gm/cm$^3$
the difference with the free-space value is about a factor of two at the center-of-mass energy of $2$~MeV.
The difference at higher gas densities are about $1-3$ orders of magnitude larger then the free-space values
at sub-barrier energies.
\begin{figure}[!htb]
	\includegraphics*[width=8.6cm]{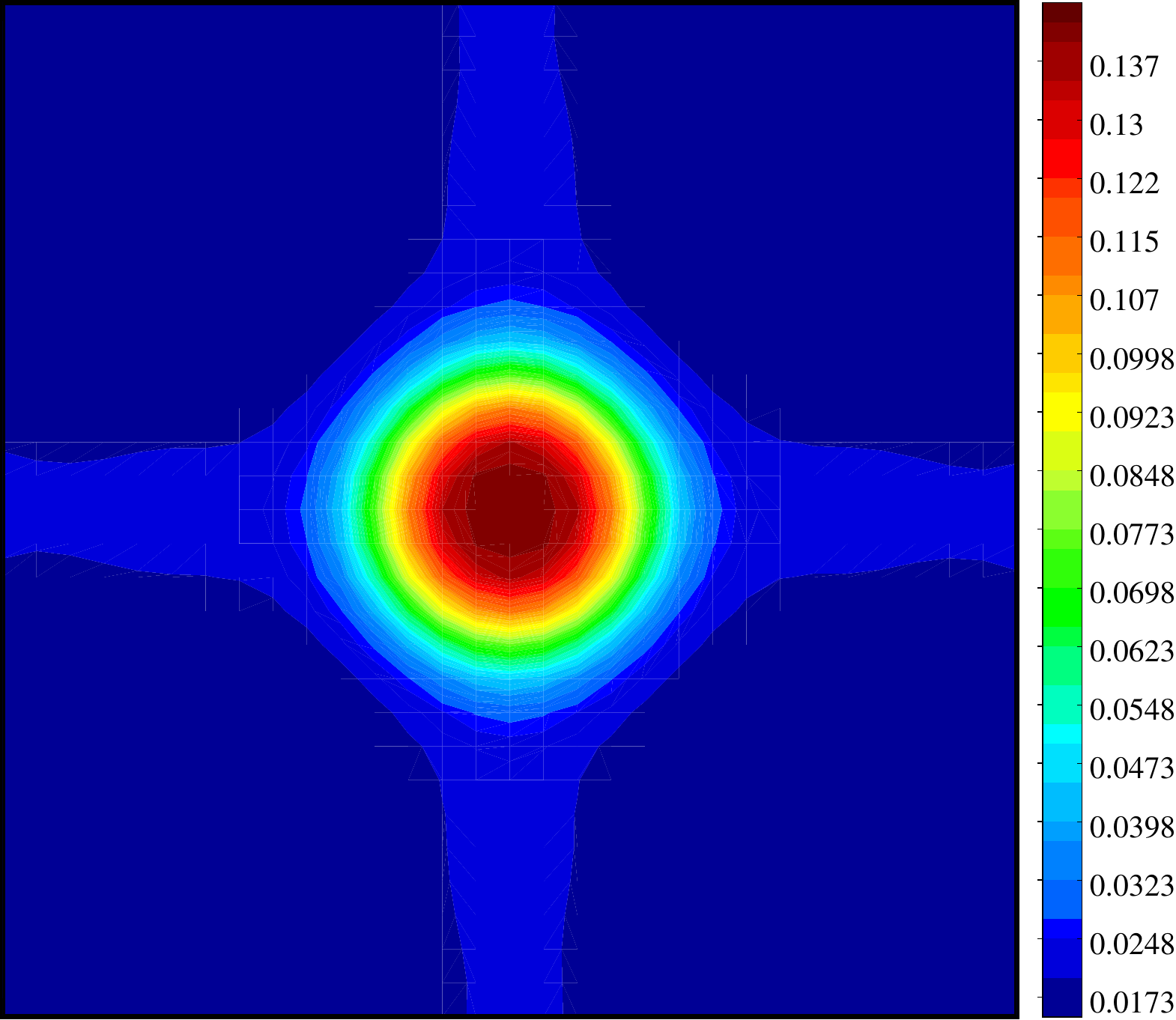}
	\caption{\protect (Color online) The cross-sectional density ($y=0$ plane) profile of the
		 $^{28}$O+500n
		system. The neutron density is low enough to have shell structures leading to the
		formation of neutron arms.
		}
	\label{fig5}
\end{figure}

In Fig.~\ref{fig5} we plot the cross-sectional density ($y=0$ plane) profile of the $^{28}$O+500n
system.
In general we see a higher gas density in the vicinity of the nucleus as can also be seen
as the dotted line in Fig.~\ref{fig2}(a).
The energy of this low density neutron gas is so low that even very small "shell effects" can lead to nonuniform densities.
In some of the cases we studied the neutron density is low enough to have shell structures
leading to the formation of neutron arms seen in Fig.~\ref{fig5}.
If we replicate this cubic box in three-dimensions one sees a lattice like structure
linked by these neutron arms.
These effects are driven mainly by the periodic boundary
conditions here. Irregularities in the grid of nuclei will
produce more irregular structures of these arms. Since
the density in the arms is of the order of free density such
that they do not affect the analysis of the bound part.

\subsection{$Z=20$ system}

\begin{figure}[!htb]
	\includegraphics*[width=8.6cm]{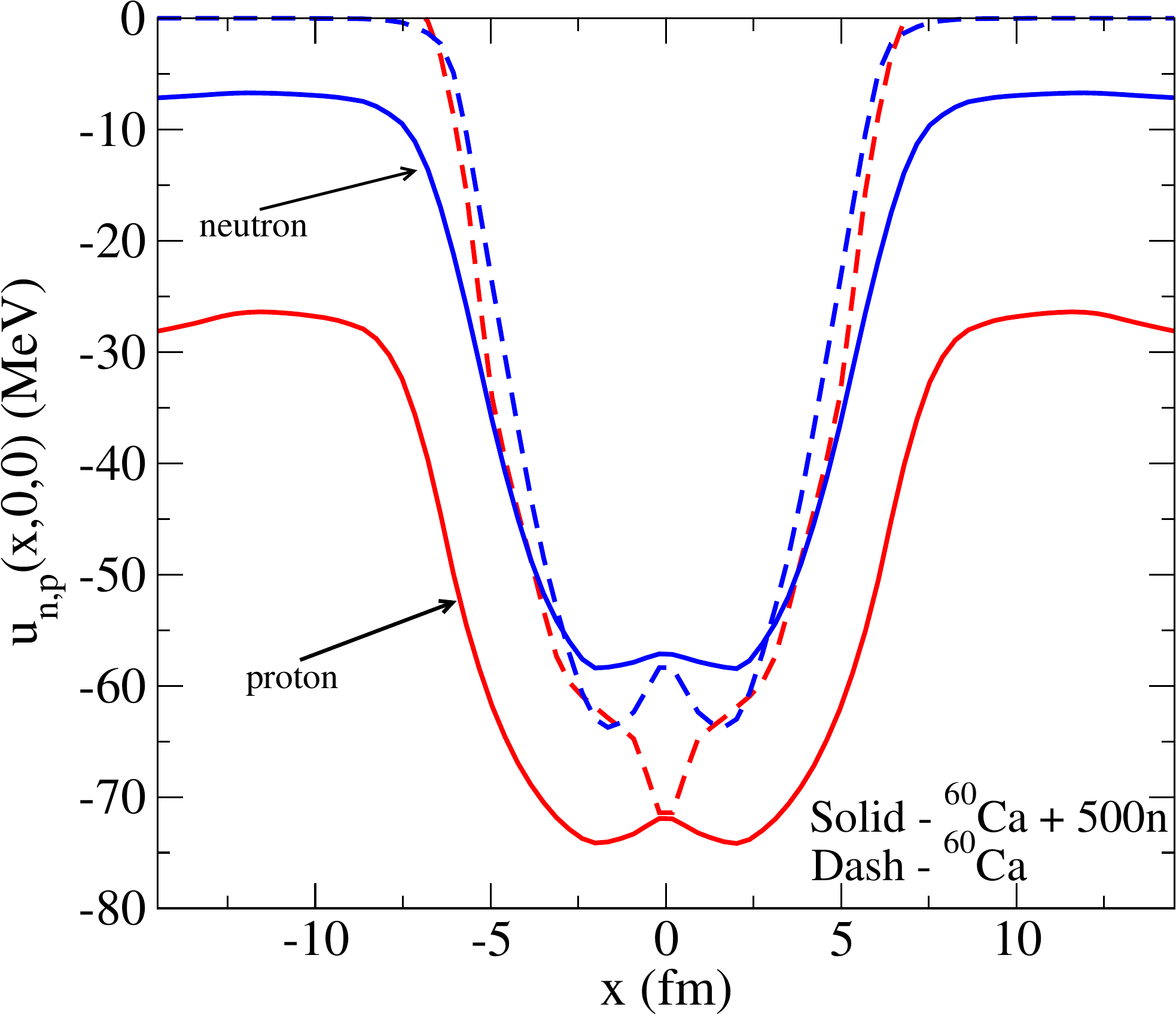}
	\caption{\protect (Color online) Mean-field potentials for neutrons and protons (solid lines) for the
		system with $Z=20$ and $540$ neutrons. The dashed lines show the mean-field
		potential of $^{60}$Ca in free-space.}
	\label{fig6}
\end{figure}

We have repeated the same study by starting with $20$ proton states and a number of neutron states
ranging from $140-1040$. Using the Skyrme SLy4 we get $^{60}$Ca as the slightly bound
drip-line nucleus in free-space. For the above range of neutron states we also find the
$^{60}$Ca to be the bound part of the system. As the number of neutron states is increased
an overall negative potential is developed permeating the entire box.
Figure~\ref{fig6} shows the neutron and proton mean-field potentials (solid lines) for
$Z=20$ and $540$ neutrons. The dashed lines show the mean-field potentials in free-space.
In this case the asymptotic value of the neutron potential is about $-6.5$~MeV.
At higher neutron densities we do observe the tendency for $^{72}$Ca to be the drip-line
nucleus but the tendency is very weak and for practical purposes considering $^{60}$Ca is sufficient for the general purposes of this study.

\begin{figure}[!htb]
	\includegraphics*[width=8.6cm]{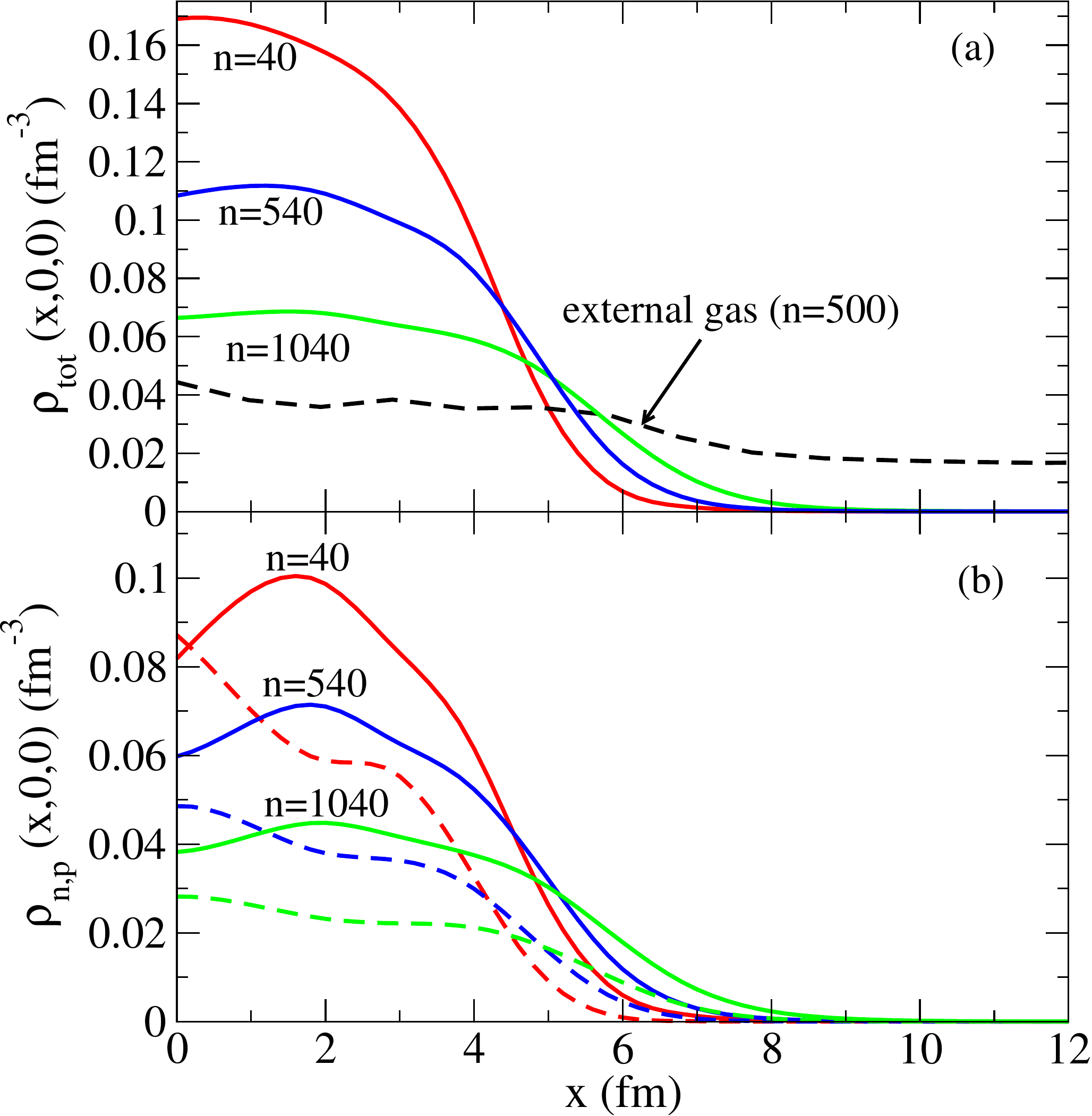}
	\caption{\protect (Color online) (a) Total density profiles for bound states; (b) density profiles for bound neutrons
		(solid lines) and protons (dashed lines), for the system with $Z=20$ and $n=40,540$, and $1040$ neutrons.}
	\label{fig7}
\end{figure}

\begin{figure}[!htb]
	\includegraphics*[width=8.6cm]{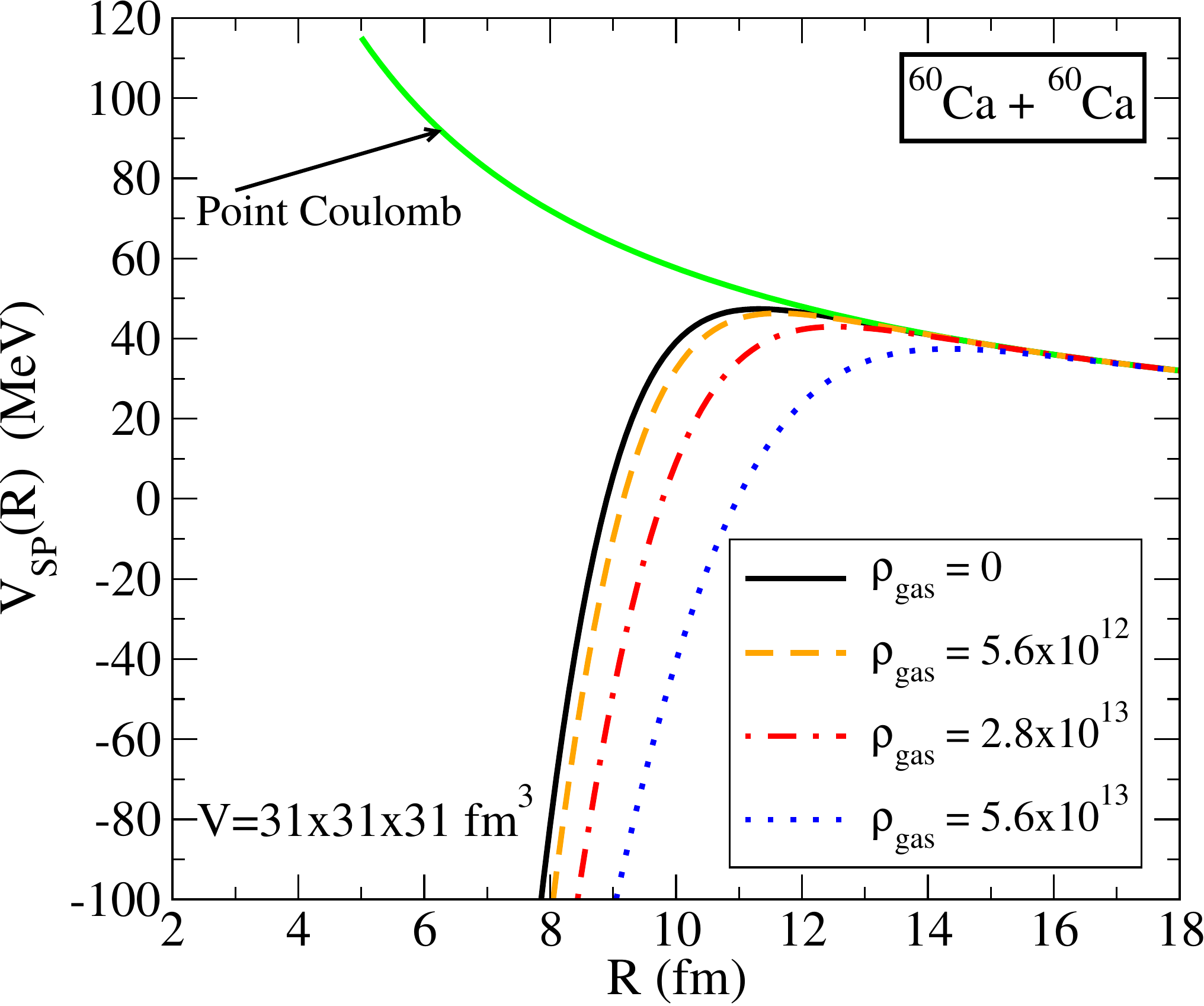}
	\caption{\protect (Color online) Ion-Ion potentials $V_{SP}(R)=V_F(R)+V_C(R)$ between two
		 $^{60}$Ca isotopes
		as a function of the external neutron-gas density. Also shown is the point-Coulomb interaction.
		Densities are in units of gm/cm$^3$.}
	\label{fig8}
\end{figure}

\begin{figure}[!htb]
	\includegraphics*[width=8.6cm]{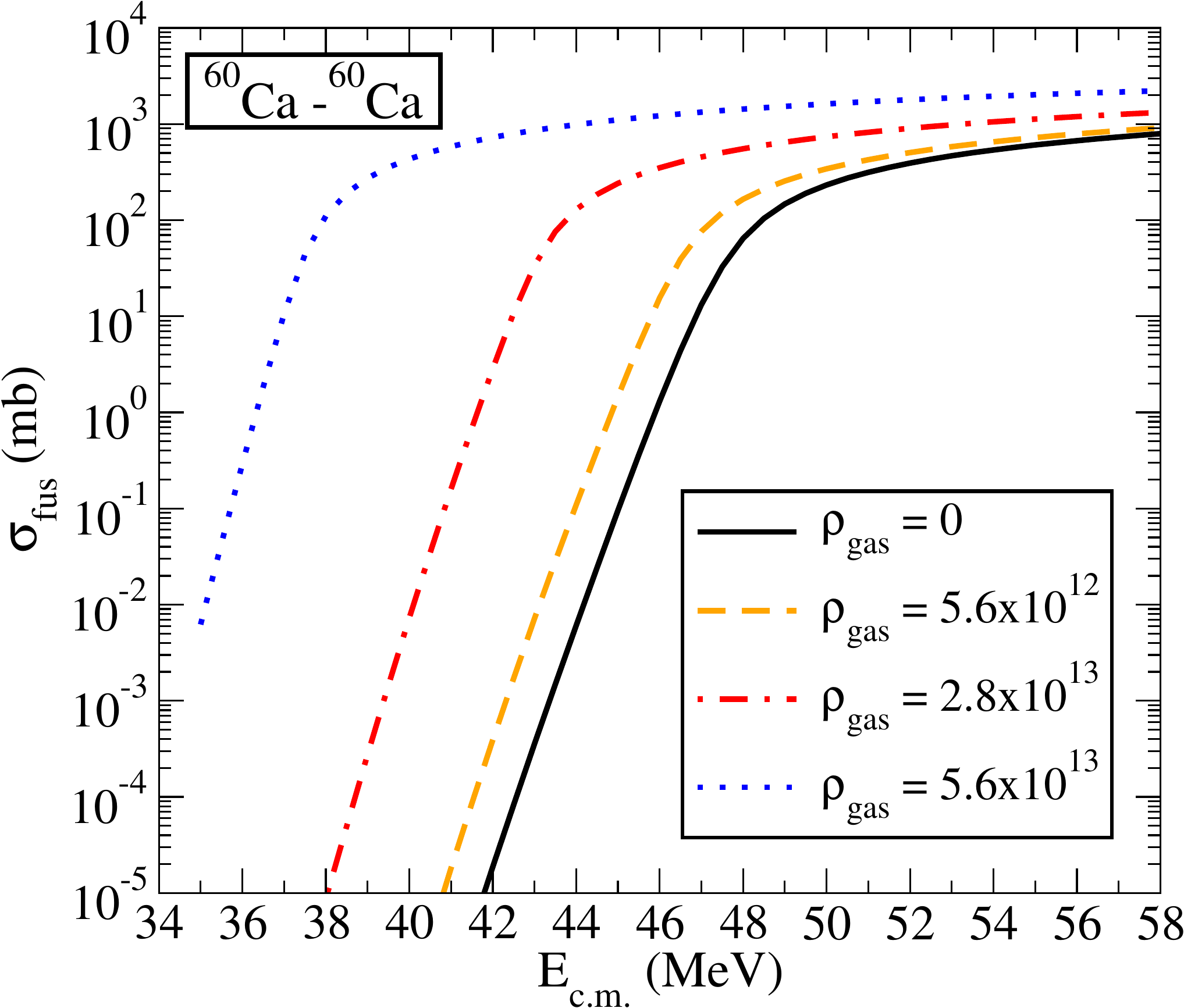}
	\caption{\protect (Color online) Fusion cross-section as a function of center-of-mass energy
		for the fusion of two $^{60}$Ca isotopes calculated for changing
		external neutron-gas density. Cross-sections are calculated using the S\~{a}o Paulo barriers
		and the IWBC method.}
	\label{fig9}
\end{figure}

In Fig.~\ref{fig7} we plot the density profiles for neutrons and protons as well as the total density
as a function of the number of neutron-gas states. The top frame shows the total density behavior
as the neutron-gas density is increased. The curves labeled $n=40$ correspond to free-space $^{60}$Ca
nucleus. As the external neutron-gas density is increased the bound system swells
up as in the $Z=8$ case.
While the peak of the total density decreases from the free-space value of $0.165$~fm$^{-3}$ to
as low as $0.048$~fm$^{-3}$ for the $1000$ external neutron-gas state case, the tail region flattens
and develops a larger spatial extent, since the total integral remains to be $60$.
The density profiles are symmetric about $x=0$ and the numerical box extends to larger values then
shown in the figure.
Corresponding ion-ion folding potentials
calculated by using these densities in Eq.~(\ref{eq:vf}) are plotted in
Fig.~\ref{fig8} for a range of external neutron-gas densities.
Again what we observe is that for neutron-gas densities in
the range $\rho_{gas}=2-4\times 10^{12}$~gm/cm$^3$ the effect of the gas in changing the
ion-ion potential compared to the free-space case is not significant. However, for gas densities
above $10^{13}$~gm/cm$^3$ a very significant change is observed.
The free-space barrier has a peak value of $47.4$~MeV located at $R=11.3$~fm. As the external gas
density is increased the corresponding barrier height is reduced to $46.3$, $42.9$, and $37.4$~MeV with the peak
location moving outward at $11.6$, $12.5$, and $14.3$~fm.
Figure~\ref{fig9} shows the fusion cross-sections calculated using the ion-ion potentials shown
in Fig.~\ref{fig8}. The dramatic rise of the cross-section is obvious as the neutron gas density
becomes significantly higher than the minimum neutron drip density.

\section{Summary and Discussion\label{sec4}}

Pycnonuclear reactions are expected to occur between very neutron-rich nuclei and in a
dense background of of a neutron gas~\cite{afanasjev2012}.
We have made an exploratory study of the effect of the external neutron gas on nuclear fusion
in the regime between the start of the neutron drip region and the melting region. 
For our study we have used an approach that treats the nuclei and the extra neutrons in a 
unified manner. The computations are done in full 3D with periodic boundary conditions and
a special treatment of the Coulomb potential with periodic boundary conditions.
For the calculation of fusion cross-sections we have used the S\~{a}o-Paulo model.
In our calculations we observe that
for lower background densities the cross-sections do not change in a very significant manner.
On the other hand as we increase the neutron gas density we observe the swelling of the nuclei
that results in the lowering of the ion-ion potential barriers and significant increase in the
fusion cross-sections.
At densities higher than the neutron-drip regime the melting of the nuclei can be observed.

While our methods give us a good understanding of fusion under these conditions more precise
computations, including the full effects of pairing and effective interactions tailored for
neutron star crust~\cite{erler2013},
may reduce some of the observed shell
effects and modify some of the results but the main features observed are expected to remain the same.
In addition, movement of nuclei inside the neutron gas as they approach each
other may cause ripples and waves in the neutron background that can also
influence these results. However, in the adiabatic limit these effects are
not expected to be very large.


\section*{Acknowledgments}
This work was supported in part by DOE grant Nos. DE-FG02-87ER40365 and DE-FG02-96ER40975,
and by the German BMBF under contract No. 05P12RFFTG.

\bibliography{VU_bibtex_master.bib}

\end{document}